\documentclass[sn-mathphys-num]{sn-jnl}

\usepackage{graphicx}%
\usepackage{multirow}%
\usepackage{amsmath,amssymb,amsfonts}%
\usepackage{amsthm}%
\usepackage{mathrsfs}%
\usepackage[title]{appendix}%
\usepackage{xcolor}%
\usepackage{textcomp}%
\usepackage{manyfoot}%
\usepackage{booktabs}%
\usepackage{algorithm}%
\usepackage{algpseudocode}%
\usepackage{listings}%
\usepackage{soul}
\usepackage{multirow}
\usepackage{booktabs}
\usepackage{colortbl}
\usepackage{amsfonts}
\usepackage{amsmath, amssymb}
\usepackage{amsthm}
\usepackage{array}
\usepackage{bbm, dsfont}
\usepackage{caption}
\usepackage{color}
\usepackage{colortbl}
\usepackage{diagbox}
\usepackage{enumitem}
\usepackage{fancyhdr}
\usepackage{float}
\usepackage{graphicx}
\usepackage{hyperref}
\hypersetup{colorlinks=true, linkcolor=blue, filecolor=black, urlcolor=blue, citecolor=blue}
\usepackage{indentfirst}
\usepackage{listings}
\usepackage{mathtools}
\usepackage{multirow}
\usepackage{multicol}
\usepackage{makecell}
\usepackage{multirow}
\usepackage{geometry}

\usepackage{algorithm}      
\usepackage{algpseudocode}  

\def\argmin{\mathop{\operator@font argmin}}
\def\argmax{\mathop{\operator@font argmax}}

\newcommand{\figlbl}[1]{\label{fig.{#1}}}
\newcommand{\figref}[1]{Fig.~\ref{fig.{#1}}}
\newcommand{\tbllbl}[1]{\label{table:{#1}}}
\newcommand{\tblref}[1]{Table~\ref{table:{#1}}}

\begin{document}

\raggedbottom

\title{Preparing for Kyber in Securing Intelligent Transportation Systems Communications: A Case Study on Fault-Enabled Chosen-Ciphertext Attacks}

\author*[1]{\fnm{Kaiyuan} \sur{Zhang}}\email{kzhang11@tufts.edu}

\author[2]{\fnm{M Sabbir} \sur{Salek}}\email{msalek@clemson.edu}

\author[3]{\fnm{Antian} \sur{Wang}}\email{antian.wang@pfw.edu}
\author[4]{\fnm{Mizanur} \sur{Rahman}}\email{mizan.rahman@ua.edu}
\author[2]{\fnm{Mashrur} \sur{Chowdhury}}\email{mac@clemson.edu}
\author[1]{\fnm{Yingjie} \sur{Lao}}\email{yingjie.lao@tufts.edu}

\affil*[1]{\orgdiv{Department of Electrical and Computing Engineering}, \orgname{Tufts University}, \orgaddress{\street{171 College Ave}, \city{Medford}, \postcode{02155}, \state{MA}, \country{USA}}}

\affil[2]{\orgdiv{Glenn Department of Civil Engineering}, \orgname{Clemson University}, \orgaddress{\street{306 S Palmetto Blvd}, \city{Clemson}, \postcode{29634}, \state{SC}, \country{USA}}}

\affil[3]{\orgdiv{Department of Electrical and Computer Engineering}, \orgname{Purdue University Fort Wayne}, \orgaddress{\street{2101 E Coliseum Blvd}, \city{Fort Wayne}, \postcode{46805}, \state{IN}, \country{USA}}}

\affil[4]{\orgdiv{Department of Civil, Construction and Environmental Engineering}, \orgname{The University of Alabama}, \orgaddress{\street{245 7th Ave Suite 3043}, \city{Tuscaloosa}, \postcode{35401}, \state{AL}, \country{USA}}}


\abstract{Intelligent transportation systems (ITS) are characterized by wired or wireless communication among different entities, such as vehicles, roadside infrastructure, and traffic management infrastructure. These communications demand different levels of security, depending on how sensitive the data is. The national ITS reference architecture (ARC-IT) defines three security levels, i.e., high, moderate, and low-security levels, based on the different security requirements of ITS applications. In this study, we present a generalized approach to secure ITS communications using a standardized key encapsulation mechanism, known as Kyber, designed for post-quantum cryptography (PQC). We modified the encryption and decryption systems for ITS communications while mapping the security levels of ITS applications to the three versions of Kyber, i.e., Kyber-512, Kyber-768, and Kyber-1024. Then, we conducted a case study using a benchmark fault-enabled chosen-ciphertext attack to evaluate the security provided by the different Kyber versions. The encryption and decryption times observed for different Kyber security levels and the total number of iterations required to recover the secret key using the chosen-ciphertext attack are presented. Our analyses show that higher security levels increase the time required for a successful attack, with Kyber-512 being breached in 183 seconds, Kyber-768 in 337 seconds, and Kyber-1024 in 615 seconds. In addition, attack time instabilities are observed for Kyber-512, 768, and 1024 under 5,000, 6,000, and 8,000 inequalities, respectively. The relationships among the different Kyber versions, and the respective attack requirements and performances underscore the ITS communication security Kyber could provide in the PQC era.}




\maketitle



\section{Introduction}
\subsection{Secure Communications in Intelligent Transportation Systems (ITS) Applications}

An intelligent transportation systems (ITS) application relies on wired and/or wireless communications among the different entities involved. For instance, connected vehicles (CVs), i.e., vehicles that communicate with other vehicles, road users, and infrastructure, are integral to ITS. CV-enabled ITS applications rely on vehicle-to-everything (V2X) connectivity to enhance transportation safety, security, mobility, and sustainability~\cite{key,qualcommCV2XAuto}. Such ITS applications may collect, aggregate, process, analyze, and disseminate information that could come from a wide array of entities. These entities can be vehicles, vulnerable and other road users, such as pedestrians and bicyclists,  transportation infrastructure, such as a traffic management center, and non-transportation infrastructure, such as a weather information service center. This information may include safety and security-related information that, if breached or tampered with, could lead to unwanted consequences, such as leakage of users' sensitive or personally identifiable information and altered vehicle safety-critical messages. Thus, maintaining the security and privacy of ITS communications at a level that is desired based on the sensitivity of communicated information is pivotal to ensuring the efficacy of the respective ITS applications. 

Utilizing appropriate encryption schemes to achieve the required level of security is essential for various communication links in an ITS application. The U.S. national ITS reference architecture (ARC-IT) assumes different security levels, such as high, moderate, and low-security levels, depending on the sensitivity of the information being transmitted \cite{ARC-IT}. For example, consider a CV speed advisory application, where the CVs receive optimal speed advisories in real-time from a roadside radio~\cite{ARC-IT}. The messages that contain speed advisory-related information might have a low-security requirement, as they are broadcast to all the CVs within the application's operational domain. On the other hand, consider an automated electronic toll collection application in which vehicle onboard devices or radios communicate with the toll collection infrastructure to pay tolls~\cite{ARC-IT}. Then, the messages sent from a CV might have a high-security requirement as the messages could include credit cards or other personally identifiable information of the users. Thus, choosing an encryption scheme that is secure enough to provide the required level of protection is vital for an ITS application. 

In addition, the beckoning of fully developed quantum computers poses a critical challenge in securing sensitive ITS communications and demands the utilization of quantum-safe encryption schemes where necessary. To this end, this study presents a general approach to applying a quantum-safe encryption scheme, i.e., Kyber, to secure communications in an ITS application. Furthermore, this study evaluates the level of security provided by the encryption scheme using a fault-enabled chosen-ciphertext attack, i.e., a benchmark attack that could exploit underlying hardware vulnerabilities to reveal the secret key of an encrypted communication channel. The next subsection presents Kyber's applicability to ITS communications and the benchmark attack considered in this study to evaluate different Kyber versions' security levels.

\subsection{Kyber for Secure ITS Communications and Potential Threats}

Kyber is a key encapsulation mechanism (KEM) designed for post-quantum cryptography (PQC), focusing on resisting attacks and other emerging threats of large-scale quantum computers. It was developed as part of the broader Cryptographic Suit for Algebraic Lattices (CRYSTALS) project. It is based on the hardness of the learning-with-errors (LWE) problem over module lattices, a problem believed to be secure against both classical and quantum computing attacks. The third round of the Public-key Encryption and Key-establishment Algorithms list was established by the National Institute of Standards and Technology (NIST) in 2020~\cite{round3}. Kyber KEM was selected as one of the finalized (i.e., almost ready for deployment) algorithms and its draft was released as part of the NIST's PQC standardization process in 2023~\cite{round3result,key2}. This highlights Kyber's potential as a foundational component in the next generation of cryptographic protocols~\cite{nistfips203,nistfips204}. 

Kyber is designed to provide security levels equivalent to the current Advanced Encryption Standard (AES) encryption system, with different parameter sets targeted at equivalent security levels of AES-128, AES-192, and AES-256, making it versatile for various security requirements. The implementation security of PQC or PQC-related algorithms has received attention recently~\cite{bao2024aeka,minhas2024edge}. The design and implementation of Kyber have seen broad integration into cryptographic libraries and systems, such as the first \textit{quantum computing safe tape drive} from IBM~\cite{ibmWorldsFirst}. From its theoretical foundations to real-world applications, Kyber has emerged as a pivotal innovation in strengthening cryptographic protocols against the rapidly evolving landscape of threats, particularly those posed by quantum computing. This progression highlights Kyber's role as a cornerstone~\cite{10323839,wang2022integral} in the ongoing effort to bolster cryptographic defenses for the future.

Kyber, now that it has been finalized as a NIST-standardized PQC algorithm, presents itself as a viable option for encrypting and decrypting ITS communications. Kyber's different versions based on different key sizes, such as Kyber-512, Kyber-768, and Kyber-1024, can be mapped to the different security levels, such as high, moderate, and low-security levels,  required by ITS applications. Thus, Kyber is a potential candidate for securing communications in different ITS applications. However, like any other encryption system, the feasibility of Kyber implementations for any ITS or related applications requires a thorough vulnerability assessment of the underlying software and hardware. This requires thorough assessments of Kyber's underlying and implementation-level vulnerabilities. For example, fault attacks could use hardware vulnerabilities by inducing errors when calculating keys. This could lead to key leakage through side-channel information. To this end, this study aims to investigate how a hardware-based attack, such as fault injections, could potentially compromise Kyber-encapsulated messages, showing the importance of robust defenses in such applications.

Fault attacks are implemented by analyzing erroneous outputs, where attackers can infer the underlying cryptographic keys or bypass security measures. Fault attacks are hazardous because they exploit physical hardware vulnerabilities, such as altering environmental conditions (such as current or voltage) or disturbing normal operations through fault injections. This type of attack underscores the need for robust hardware design and fault-tolerant cryptographic algorithms that can resist such manipulations. For example, the Clock Glitch Attack is an attack that manipulates the clock signals of a device to cause it to malfunction temporarily~\cite{9250822}. An attacker can cause errors in Rivest-Shamir-Adleman (RSA) encryption computations by overclocking an intelligent card beyond its operational specifications, potentially leading to the leakage of private keys~\cite{ravi2020drop,guo2020key,bhasin2021attacking,oder2016practical}. These time-side rotational information leakage usually could play a critical role in a time-side-channel attack. 

The fault attack we considered in this paper is a Chosen-Ciphertext Attack (CCA)~\cite{ravi2020drop,hamburg2021chosen}. CCA is an attack in which an attacker can choose a specific bit to flip in a ciphertext and generate inequalities to derive the corresponding decrypted plaintext. This capability allows the attacker to exploit vulnerabilities in a cryptographic system to decrypt other ciphertexts or even recover the secret key. The combination of fault injections and CCA poses significant threats to Kyber implementations~\cite{ravi2020drop,hamburg2021chosen}. The fault-enabled CCA involves manipulating the ciphertext and observing the system's behavior when faults are introduced during decryption. It exploits the Fujisaki-Okamoto transformation~\cite{fujisaki1999secure, hofheinz2017modular}, a common technique used in KEMs to achieve CCA security. The attacker observes whether decapsulation fails or succeeds in deriving information about the secret data. For example, Hermelink et al.~\cite{hermelink2021fault} demonstrated that even secure systems might be vulnerable against traditional CCA when hardware faults are considered. The authors in~\cite{hermelink2021fault} highlighted that while standard countermeasures such as shuffling and boolean masking could protect against some attacks, combining fault injections with chosen ciphertext attacks could overcome these defenses.

\subsection{Contribution}

In this study, we present a generalized approach for leveraging Kyber KEM to achieve the desired level of security in ITS communication. Our key contributions include implementing and analyzing fault-enabled CCA at different Kyber security levels and evaluating how these levels impact encryption and decryption performance during the attack. By mapping the security requirements of ITS applications to Kyber's security parameters, we provide a practical assessment of Kyber's resilience. The findings offer valuable insight into the relationships among the Kyber security levels, the fault attack performance, and the corresponding computational costs, guiding the choice of suitable cryptographic parameters to enable secure communications in an ITS application.

\section{Theoretical and Conceptual Background}
\subsection{CRYSTALS-Kyber }
Kyber relays a lattice-based cryptography problem: learning-with-errors (LWE), which involves solving linear equations with added random noise, making it difficult to determine the original solution. It has three parameter sets: Kyber-512, Kyber-768, and Kyber-1024. Kyber uses a public key encryption (PKE) scheme followed by a KEM derived from the Fujisaki-Okamoto transform variant \cite{fujisaki1999secure}. 

Kyber PKE: PKE.KeyGen, PKE.Encrypt, PKE.Decrypt. The simplified decapsulation process is shown in the Algorithm~\ref{algo1}.\textbf{Kyber PKE.} In Kyber, the computations are performed in the rings \( R_q \) and \( R_{q, \kappa} \), where \( R_q = \mathbb{Z}_q[x]/(x^n + 1) \), with the parameters \( n = 256 \), \( q = 3329 \),  and \( \kappa \in \{2, 3, 4\} \), determined by the specific parameter set. $\mathbb{Z}_q$ denotes the ring of integers modulo $q$ and $x$ is an indeterminate used in the polynomial ring. The polynomials in $\mathbb{Z}_q[x]$ have coefficients that are elements of $\mathbb{Z}_q$. The ring $\mathbb{R}_q$ is then defined as the quotient ring $\mathbb{Z}_q[x]/(x^n + 1)$, which means polynomials are considered equivalent if they differ by a multiple of $x^n + 1$. Sampling is deterministic and relies on an incremented seed after each call. The functions of compression and decompression are defined as follows:
\begin{equation}
   \text{Compress}_q(x,d)=\lceil\frac{2^d}{q}\cdot x\rfloor \text{ mod } 2^d\\
\end{equation}
\begin{equation}
    \text{Decompress}_q(x,d)=\lceil\frac{q}{2^d}\cdot x\rfloor \\
\end{equation}
where, $d$ is set to $d_u$ or $d_v$. For Kyber-512, we set the parameters as follows: ($\eta_1$, $\eta_2$) = (3, 2), ($d_u$, $d_v$) = (10, 4).  The functions \textbf{Encode} and \textbf{Decode} provide interpretation between a bitstream and a polynomial representation. The \textbf{Decode} function works by scaling each bit of the message \( m \) by \( q \). The function \( f \) performs rounding to the nearest integer, while the modulo operation \( \text{mod}\ q \) maps an integer \( x \) to a value \( x' \in \{0, 1, \ldots, q - 1\} \), satisfying \( x \equiv x' \pmod{q} \). Elements processed using the Number Theoretic Transform (NTT) are represented as \( \hat{a} = \text{NTT}(a) \). Vectors and matrices, corresponding to elements in \( R_q^{\kappa} \) and \( R_q^{\kappa \times \kappa} \), are denoted by bold lowercase and uppercase letters, respectively. For any \( a \in R_q^{\kappa} \), \( \text{NTT}(a) \) refers to applying the NTT to each component independently.

\begin{algorithm}[htbp]
\caption{\textbf{Kyber-KEM Decapsulation (simplified)}}\label{algo1}
\begin{itemize}
\item \textbf{Input:} Secret key $sk = (\hat{\mathbf{s}}, pk, z)$, ciphertext $ct = (\mathbf{c_1}, c_2)$
\item \textbf{Output:} Shared key $K$
\end{itemize}

\begin{enumerate}
  \item $m \leftarrow \text{PKE.Decrypt}(sk, ct)$ \hfill $\triangleright$ Retrieve seed for re-encryption
  \item $(\Bar{\mathbf{K}}, \tau) \leftarrow G(m||H(pk))$ \hfill $\triangleright$ Retrieve seed for re-encryption
  \item $c' \leftarrow \text{PKE.Encrypt}(pk, m, \tau)$ \hfill $\triangleright$ Re-encrypt
  \item \textbf{if} $c = c'$ \textbf{then} 
        \item \hspace*{6mm} $K \leftarrow \text{KDF}(\Bar{\mathbf{K}}||H(c))$ \hfill $\triangleright$ Derive shared key on successful re-encryption
  \item \textbf{else}
        \item \hspace*{6mm} $K \leftarrow \text{KDF}(z||H(c))$ \hfill $\triangleright$ Implicit rejection on failure
  \item \textbf{return} $K$
\end{enumerate}
\end{algorithm}


\subsection{Fault-Enabled Chosen-Ciphertext Attacks (CCAs) on Kyber}
In the previous work~\cite{prokop2021fault}, the attack is highly implementation-specific and requires a high level of synchronization. Also, the countermeasure is simple~\cite{prokop2021fault}, such as shuffling. Those limitations have been overcome in some relatively recent work; for example, Hermelink et al. ~\cite{hermelink2021fault} introduced an instruction-skipping fault in the Decompress/Decode method. The authors in~\cite{hermelink2021fault} proved that the Fujisaki-Okamoto transform equality check is not the only critical operation for fault attacks.

In the process of introducing a fault and subsequently verifying whether the re-encryption comparison fails---by determining if the device still produces the correct shared secret \( K \)---a linear inequality related to the decryption error polynomial is derived (refer to Equation (1)). Specifically, during the \( j \)-th fault injection, this error polynomial can be represented as:
\begin{equation}
e^T r_j - s^T(e_{1j} + \Delta u_j) + e_{2j} + \Delta v_j 
\end{equation}
where the vector \( e^T \) represents the error terms associated with the encryption process, while \( e_{1j} \) and \( e_{2j} \) are specific components of the error polynomial during the \( j \)-th fault injection. The vector \( s^T \) denotes the secret key used in the decryption process, which is essential to decrypt the ciphertext and recover the shared secret key. For \( j \in \{0, \ldots, l - 1\} \), \( l \) represents the total faults number introduced, \( \Delta u_j \) and \( \Delta v_j \) correspond to the compression errors in \( u_j \) and \( v_j \), respectively, through the application of compression and decompression. It is important to note that the values \( r, e_1, e_2, \Delta u, \Delta v \), along with the true shared secret \( K \), are all assumed to be known to the attacker, under the assumption that the attacker faithfully follows the encapsulation process. Representing the $t$-th component of a vector of polynomials by $r^{(t)}$, the $i$-th coefficient of the error term polynomial is given by,

\begin{equation}
\begin{split}
&\sum_{t=0}^{k-1} \sum_{h=0}^{n} \sigma(h, i)e_{h^{(t)}},r_{j,\tau(h,i)}^{(t)} \\ 
&+\sum_{t=0}^{k-1} \sum_{h=0}^{n} \sigma(h, i)s_{h^{(t)}}(e1_{j,\tau(h,i)}^{(t)} + \Delta u_{j,r_{j,(h,i)}}^{(t)} )\\
&+e2_{j,i} + \Delta v_{j,i,}
\end{split}
\end{equation}
where, $\sigma(h,i) = i - h \mod n$, and $\sigma(h,i)$ returns 1 if $i - h \geq 0$ and $-1$ otherwise. The notation $(\cdot)_i$ represents all associated sign inversions and index transformations, and the above expression can be equivalently reformulated in terms of dot products.

\begin{equation}
(r_j, e) + ((e_{1j} + \Delta u_j), s) + e_{2j,i} + \Delta v_{j,i}
\end{equation}
where each coefficient of the polynomials \( e \) and \( s \) is sampled from a known binomial distribution with limited support. To recover the secret key from the previously mentioned inequalities, Pessl~\cite{prokop2021fault} proposed initializing a probability distribution vector of length \( 2n \) using this binomial distribution. This vector is iteratively refined based on the information obtained from the system of inequalities represented by matrix \( A \) and vector \( \mathbf{b} \). As noted earlier, the attack depends on the specific implementation and requires precise synchronization, particularly skipping over a specific instruction. Moreover, this type of attack can likely be mitigated through simple techniques such as code shuffling. Therefore, the full potential of fault attacks remains uncertain and is open for further investigation.

The ciphertext in a PKE scheme consists of the compressed polynomial \( v \), resulting in \( c_2 \), and the compressed vector of polynomials \( u \), which produces \( c_1 \). During decryption, the function decompresses both \( c_1 \) and \( c_2 \), recovering approximate versions of the polynomials \( u \) and \( v \). It then computes an approximate version of \( v - u^T s \), expressed as:
\begin{equation}
\text{rec} = e^Tr - s^T(e_1 + \Delta u) + e_2 + \Delta v + \text{Decode}(m)
\end{equation}
Here, each coefficient is reduced to the range \( \{0, \ldots, q-1\} \), where the \( \Delta \)-terms represent the error introduced by the compression and the subsequent decompression of the polynomial(s) or vector of polynomials. The message is recovered by mapping the coefficients of \( \text{rec} \), specifically \( \text{rec}_i \) for \( i \in \{0, \ldots, n - 1\} \), to a 0-bit if \( \text{rec}_i \) is closer to 0 or \( q \) than to \( q/2 \). Otherwise, \( \text{rec}_i \) is assigned to 1. Thus, the function that maps a coefficient reduced to \( \{0, \ldots, q-1\} \) to a bit is given by:
\begin{align}
    \phi& : \{0, \ldots, q - 1\} \rightarrow \{0, 1\}\\
    a &\mapsto 
\begin{cases} 
0, & \text{if } \min(|a - q|, |a|) < q/4 \\
1, & \text{else.}
\end{cases}
\end{align}
This approach yields the message \( m \) with high probability for a given ciphertext, as the error polynomial \( d \), defined as follows, remains small.
\begin{align}
    d = e^Tr - s^T(e_1 + \Delta u) + e_2 + \Delta v
\end{align}







The fault-enabled CCA considered in this study is carried out by (i) manipulating and correcting the ciphertext and (ii) obtaining inequalities and recovering the secret key using belief propagation techniques. The following subsections detail the attack process:
\subsubsection{Manipulating and Correcting the Ciphertext}
The attack begins with the manipulation of a valid ciphertext by flipping a single bit in it. The modified ciphertext is then used to attempt decryption. During decryption, a fault is deliberately introduced to correct the manipulated ciphertext by flipping the altered bit back to its original state. The Fujisaki-Okamoto transformation, integrated within the decryption process, is exploited as an oracle. The attacker can infer certain secret key properties by observing whether the decryption succeeds or fails.

\subsubsection{Obtaining Inequalities and Recovering the Secret Key}
Linear inequalities involving the secret key are derived from the outcomes of the decryption process whether it succeeds or fails—following bit manipulation and fault injection. \figref{Manipulating_and_correcting_the_ciphertext} shows the ciphertext manipulating processing and the attack position.  The set of inequalities is gradually constructed by repeating this procedure with different bits and fault injections. Each distinct manipulation generates a new inequality, thereby enriching the dataset from which the secret key can be inferred. Recovering the secret key from these inequalities in a fault-enabled CCA employs belief propagation techniques. In this approach, messages received at each variable node are used to update the probability distributions associated with the corresponding variables. This step involves convolving probability distributions to incorporate new information from the check nodes. The final reconstruction of the secret key is achieved by combining the estimated coefficients following the cryptographic scheme's key structure. Additional procedures, such as solving linear systems or conducting cryptographic validation, may be necessary to verify the accuracy of the recovered key.

\begin{figure}[htbp]
    \centering
     \includegraphics[width=0.9\textwidth]{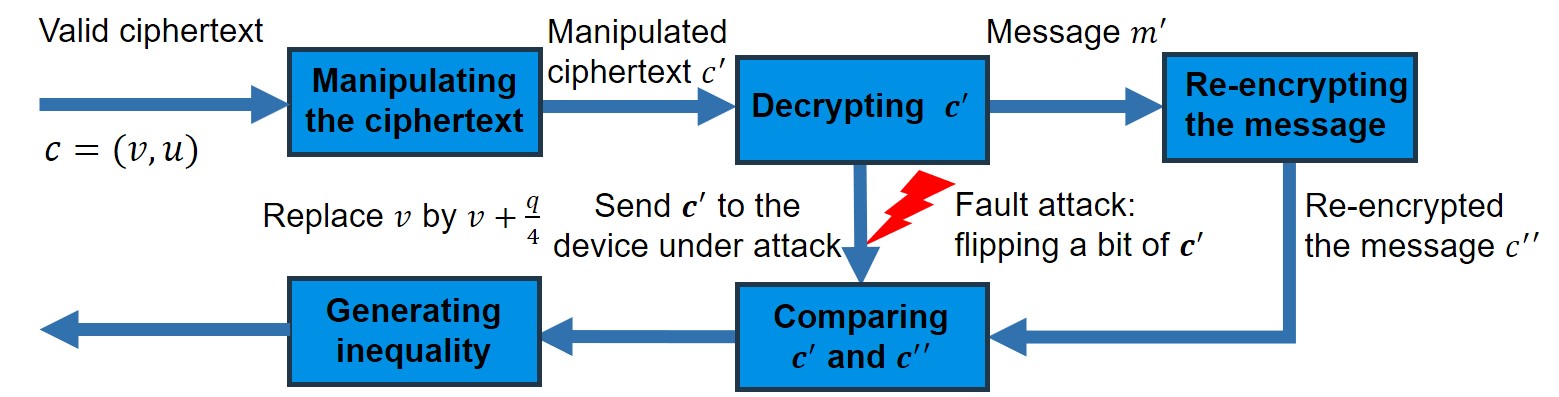}
    \caption{Manipulating and correcting the ciphertext}
    \figlbl{Manipulating_and_correcting_the_ciphertext}
\end{figure}

\section{Encryption Using Kyber and Considered Attack Configurations}
\subsection{Encryption Using Kyber}
To explore the feasibility of Kyber implementation for encrypting messages in ITS communication, we used a simplified version of Kyber called the \textit{baby Kyber}~\cite{cryptopediaKyberDoes}. This version maintains the same structure as the Kyber-512 algorithm but uses reduced polynomial parameters instead of the original ones. Table~\ref{Baby_Kyber_parameters} provides an overview of the parameters used for the \textit{baby Kyber} algorithm. In contrast, for the full Kyber implementation, the tests in this paper are based on the Kyber algorithm, as standardized by NIST \cite{bos2018crystals}. 

\begin{table}[htbp]
\centering
\caption{Baby Kyber parameters}\label{Baby_Kyber_parameters}

\begin{tabular}{cccccc}
\hline
$q = 17$ (plain modulus) \\ \hline
$f = [1,0,0,0,1]$ (poly modulus, $x^4+1$)           \\ \hline
$s = [[0, 1, -1, -1], [0, -1, 0, -1]]$  (secret key: $-x^3-x^2+x, -x^3-x]$)  \\ \hline
$A = [[[11, 15, 14, 6], [3, 6, 7, 9]], [[12, 10, 3, 5], [15, 4, 1, 9]]]$ (public key)             \\ \hline
$e = [[0, 0, 1, 0], [0, -1, 1, 0]]$ (noise)          \\ \hline
$r = [[0, 1, 0, -1], [-1, 1, 0, 1]]$ (blinding vector for encrypt)         \\ \hline
$e_1 = [[1, 0, 1, 0], [0, 0, 1, 0]]$ (noise vector for encrypt) \\ \hline
$e_2 = [0, -1, 0, -1]$ (noise vector for encrypt) \\ \hline 
\end{tabular}
\end{table}

We used three security levels of Kyber, as detailed in Table~\ref{Default_experiment_setting}. Additionally, we implemented a fault-enabled CCA~\cite{hermelink2021fault} on each Kyber implementation to determine an appropriate attack time cost for each security level. All experiments were conducted on an Intel(R) Xeon(R) Gold 6346 CPU, utilizing 64 threads for each test.

\subsection{Fault-Enabled CCA Configurations}
We assume that the adversary can intercept and capture ciphertexts and possesses preliminary knowledge of the Kyber algorithm, which enables a fault-enabled CCA. In this paper, the fault attack utilizes belief propagation to recover the secret key by solving a linear system of inequalities. Belief propagation is applied through variable nodes and check nodes. The variable nodes are initialized with the binomial distribution associated with the error and the secret key. The check nodes receive distributions from the variable nodes, update the distributions, and compute the probability distribution for each key coefficient. By controlling the number of inequalities, our fault-enabled CCA aims to identify an optimal time-power trade-off for the attack.


\begin{table}[htbp]
\centering
\caption{Default experiment setting}\label{Default_experiment_setting}
\begin{tabular}{cccccc}
\hline
Iteration & Threads & Security level & Text length \\ \hline
100       & 64      & 512            & 1024        \\ \hline
100       & 64      & 768            & 1024        \\ \hline
100       & 64      & 1024           & 1024        \\ \hline
\end{tabular}

\end{table}

\begin{figure}[htbp]
    \centering
    \resizebox{0.7\textwidth}{!}
     { \includegraphics{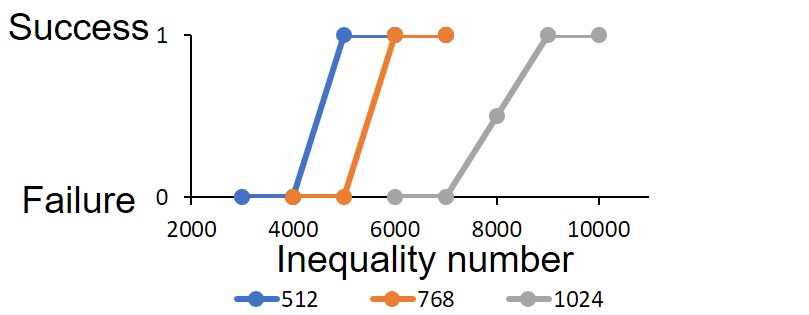}}   
    \caption{Results from Attack Experiments on Baby Kyber Implementation (0 represents attack failure and 1 represents attack success).}
    \figlbl{Default Experiment result}
\end{figure}

\begin{figure}[htbp]
    \centering
    \resizebox{0.7\textwidth}{!}
     { \includegraphics{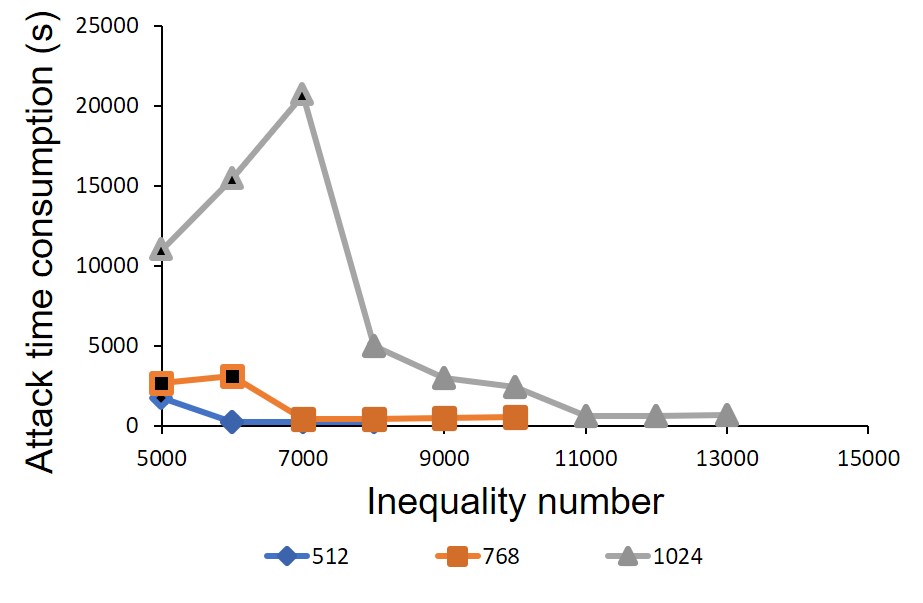}}   
    \caption{Breach time for three Kyber levels with inequalities number 5,000-13,000 (The black dots represent failure attacks)}
    \figlbl{Time experiment 1}
\end{figure}

\if false

\begin{table}[htbp]
\centering
\begin{threeparttable}
    \caption{Percentage error}
    \begin{tabular}{cccccc}
    \hline
Percentage & 5000   & 6000    & 7000    & 8000    \\ \hline
512        & 5.96\% & 8.81\%  & 14.84\% & 8.61\%  \\ \hline
768        & 6.29\% & 65.29\% & 4.92\%  & 5.91\%  \\ \hline
1024       & 0.46\% & 0.50\%  & 0.43\%  & 20.76\% \\ \hline
    \end{tabular}
    \tbllbl{Percentage error}
\begin{tablenotes}
\item[$\dagger$] Percentage error for each setting
\end{tablenotes}
\end{threeparttable}
\end{table}

\fi

\section{Evaluation Using an ITS Use Case}

This section presents the experimental evaluation of Kyber-encapsulated encrypted communication with an ITS use case. Section 4.1 presents the ITS data considered for encryption in this study. Sections 4.2 through 4.5 present the evaluation results of our experiments with fault-enabled CCA to reveal the secret key in a Kyber-encapsulated ITS communication. 

\subsection{ITS Use Case and Data} 
We considered Kyber for encrypting messages from CVs in an ITS application. The messages are encrypted using Kyber on the host side, i.e., a CV that transmits the messages. The CV sends the ciphertext to the receiver(s), such as a receiving CV or a roadside infrastructure. The receiver(s) then decrypt the ciphertext on the receiver side.

CVs utilize V2X communication technology to communicate with other vehicles, roadway users, and roadside infrastructure~\cite{key,qualcommCV2XAuto}. Onboard wireless communication radios enable CVs to establish message-based V2X communication with various entities. The SAE J2735 standardizes the structure of the V2X messages for any ITS applications: V2X Communications Message Set Dictionary ~\cite{j2735}. These messages can include different types of information depending on the respective ITS application type; for example, message ID, transponder ID, vehicle class, heading, latitude, longitude, altitude, and vehicle basic and control statuses. Based on the message contents, different communication links in a CV environment may have different security requirements, such as high, moderate, and low-security requirements. For example, in a CV-based electronic toll collection system~\cite{9344812}, the vehicles may need to encrypt sensitive or personally identifiable messages. This information can be related to the vehicles, such as transponder ID, vehicle license plate number, and vehicle identification numbers, or the information can be related to the users, such as social security and credit card information. Another example in which CV messages might be required to be encrypted is a vehicular social network (VSN)~\cite{deng2020location}. A vehicular social network is a special type of vehicular ad hoc network (VANET), in which protecting the privacy of location information (e.g., latitude, longitude, and altitude) using some suitable encryption scheme is often crucial. In this study, we chose Kyber to ensure the security of the shared secret information in V2X messages, structured following the SAE J2735 standard.

The V2X data used in this study is the core data of a real-world CV basic safety message (BSM). The Tampa Hillsborough Expressway Authority (THEA) CV Pilot program ~\cite{BSM} collected the data utilized in this study. Table~\ref{MSB_content_description} explains the data elements of the BSM core data used in this study. The BSM core data is a sequence of safety data related to a vehicle's state that is always included in every BSM~\cite{j2735}. Like a key-value pair data structure, the V2X or BSM core data used in this study includes a value corresponding to each data element presented in Table~\ref{MSB_content_description}. Although the experiments conducted in this study considered V2X data of the format shown in Table~\ref{MSB_content_description}, it is simply a plaintext for the Kyber encryption scheme. Once encrypted, the plaintext becomes a string of numbers, i.e., a ciphertext with no contextual meaning until we decrypt it. Thus, changing a plaintext, for instance, using a different V2X data, is not expected to affect the experiments we conducted. This implies that the results presented in the following subsections are generalizable for any ITS or non-ITS-related messages.

\begin{table}[htbp]
\centering
\caption{Data Elements of the BSM Used in This Study (adopted from~\cite{BSM})}
\begin{tabular}{|>{\centering\arraybackslash}m{3cm}|p{11cm}|}
\hline
\textbf{Content} & \textbf{Description} \\ \hline
msgCnt           & A sequence number within a stream of messages from the same sender \\ \hline
id               & A random device identifier which changes periodically to ensure anonymity \\ \hline
secmark          & A second mark that represents the milliseconds within a minute \\ \hline
lat              & Geographic latitude of the vehicle \\ \hline
long             & Geographic longitude of the vehicle \\ \hline
elev             & Geographic position above or below the reference ellipsoid (typically based on the World Geodetic System 1984) \\ \hline
\multirow{1}{*}{accuracy} 
                 & semiMajor: Positional accuracy of the semi-major axis of an ellipsoid        representing the expected accuracy from a GNSS; \newline
                   semiMinor: Positional accuracy of the semi-minor axis of an ellipsoid representing the expected accuracy from a GNSS; \newline
                   orientation: Positional accuracy of the semi-major axis orientation of an ellipsoid representing the expected accuracy from a GNSS with respect to the coordinate system \\ \hline
transmission     & The current state of the vehicle transmission (e.g., neutral, park, forwardGears, reverseGears, unavailable) \\ \hline
speed            & The vehicle speed \\ \hline
heading          & The heading angle of the vehicle \\ \hline
angle            & The steering wheel angle \\ \hline
\multirow{4}{*}{accelSet} 
                 & long: Acceleration value along the vehicle's longitudinal axis; \newline
                   lat: Acceleration value along the vehicle's lateral axis; \newline
                   vert: Acceleration value along the vehicle's vertical axis; \newline
                   yaw: Yaw rotation rate of the vehicle about its vertical axis \\ \hline
\multirow{6}{*}{brakes} 
                 & wheelBrakes: The wheel brake applied status; \newline
                   traction: The status of the vehicle traction control system; \newline
                   abs: The status of the vehicle antilock braking system (ABS); \newline
                   Scs: The status of the vehicle stability control system (SCS); \newline
                   brakeBoost: The status of auxiliary brakes; \newline
                   auxbrakes: The status of auxiliary brakes \\ \hline
\multirow{2}{*}{size} 
                 & width: Width of the vehicle; \newline
                   length: Length of the vehicle \\ \hline
\end{tabular}
\label{MSB_content_description}
\end{table}

\subsection{Inequalities Required to Recover the Secret Key} 
In this study, we considered encrypting a sample V2X message from using different Kyber versions associated with different security requirements. Initially, we use the default parameters settings of \textit{baby Kyber}, shown in Table~\ref{Default_experiment_setting}, to experiment. To determine the approximate threshold for the number of inequalities where the attack switches between success and failure for each Kyber security level, we conducted fault-enabled CCA experiments on Kyber-512, Kyber-768, and Kyber-1024. The inequalities ranged from 3,000 to 6,000 for Kyber-512, 4,000 to 7000 for Kyber-768, and 6,000 to 10,000 for Kyber-1024. The results are presented in \figref{Default Experiment result}. It is observed from \figref{Default Experiment result} that 5,000, 6,000, and 9,000 inequalities are required to achieve attack successes under the default parameter settings of Kyber-512, Kyber-768, and Kyber-1024 versions. This finding aligns with the increased security level provided by higher Kyber versions.

\subsection{Attack Time Consumption} 
In the previous subsection, we used the \textit{baby Kyber} for V2X message encapsulation to test our workflow's correctness and to determine the approximate number of inequalities required to recover the secret key. This subsection moves to experiment with the full Kyber implementation (as standardized by NIST~\cite{bos2018crystals}).

Different Kyber security levels require varying numbers of inequalities to recover the secret key, as observed from \figref{Default Experiment result} for the \textit{baby Kyber} parameter settings. Consequently, the time required to implement a fault-enabled CCA varies across different security levels. Table~\ref{Time_experiment_1_tb} and \figref{Time experiment 1} present the time taken for the attack under different inequality number conditions. In Table~\ref{Time_experiment_1_tb}, the red cells indicate failed attacks, while the green cells represent successful attacks. From \figref{Time experiment 1}, it is observed that Kyber-1024 is the most secure among the three versions of Kyber we considered in this study and the most time-consuming. For Kyber-512 and Kyber-768, when the attack is successful, the time consumption increases gradually with the increased number of inequalities, as observed from Table~\ref{Time_experiment_1_tb}. On the other hand, for Kyber-1024, once the attack is successful, the time decreases as the number of inequalities increases. However, beyond 11,000 inequalities, the time consumption is slowly increasing for Kyber-1024. For Kyber-1024, adding more inequalities initially helps by narrowing down the solution space, making the attack more efficient. However, after a certain point, the increased number of inequalities adds too much complexity and redundant information, which slows down the attack instead of helping it.

\begin{figure}[htbp]
    \centering
    \resizebox{0.7\textwidth}{!}
     { \includegraphics{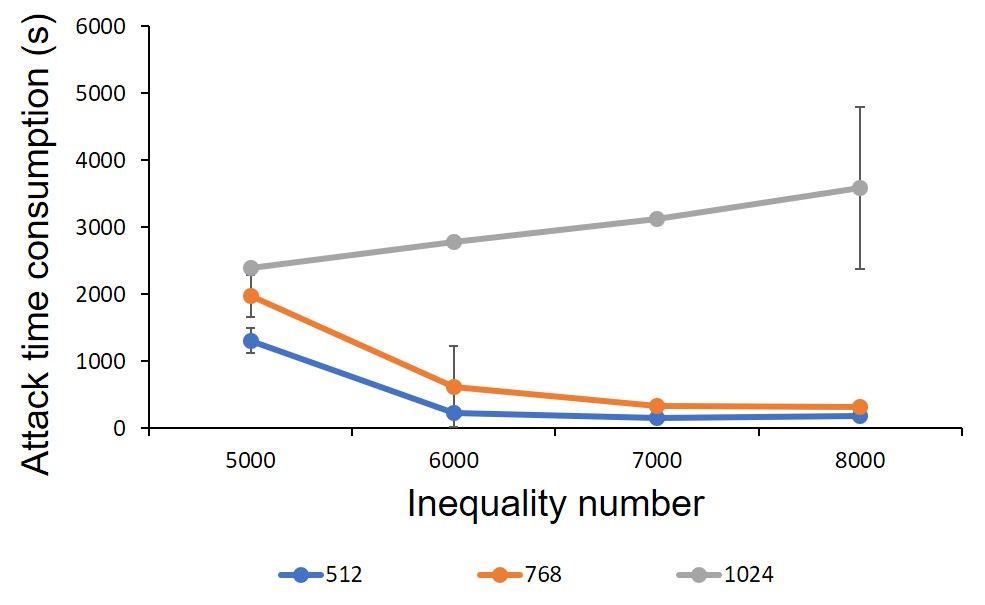}}   
    \caption{Average breach time for three Kyber levels with inequalities number 5,000-8,000 in five times repeat CCA experiment.}
    \figlbl{Time experiment 2}
\end{figure}

\begin{table}[htbp]
\centering
\caption{Breach Time for Three Kyber Levels with Different Inequality Numbers}
\label{Time_experiment_1_tb}
\begin{tabular}{|c|c|c|c|c|}
\hline
\textbf{Kyber Version} & \textbf{5000} & \textbf{6000} & \textbf{7000} & \textbf{8000} \\ \hline
512 &
  \cellcolor[HTML]{FF0000}\textbf{1728.4} &
  \cellcolor[HTML]{92D050}252.2 &
  \cellcolor[HTML]{92D050}259.5 &
  \cellcolor[HTML]{92D050}272.7 \\ \hline
768 &
  \cellcolor[HTML]{FF0000}\textbf{2659.3} &
  \cellcolor[HTML]{FF0000}\textbf{3112.4} &
  \cellcolor[HTML]{92D050}453.8 &
  \cellcolor[HTML]{92D050}439.2 \\ \hline
1024 &
  \cellcolor[HTML]{FF0000}\textbf{10991} &
  \cellcolor[HTML]{FF0000}\textbf{15444.8} &
  \cellcolor[HTML]{FF0000}\textbf{20698.9} &
  \cellcolor[HTML]{92D050}5022.2 \\ \hline
\textbf{} & \textbf{9000} & \textbf{10000} & \textbf{11000} & \textbf{} \\ \hline
512 & \#N/A & \#N/A & \#N/A & \\ \hline
768 &
  \cellcolor[HTML]{92D050}498.2 &
  \cellcolor[HTML]{92D050}540.2 &
  \#N/A &
  \\ \hline
1024 &
  \cellcolor[HTML]{92D050}2971.2 &
  \cellcolor[HTML]{92D050}2404.8 &
  \cellcolor[HTML]{92D050}615.67 &
  \\ \hline
\end{tabular}
\footnotetext{The time is presented in seconds. Red cells represent attack failure. Green cells represent attack success.}
\end{table}

\subsection{Attack Time Instability} 
In certain conditions, the fault-enabled CCA demonstrates instability. We repeated the fault attack five times with inequality numbers ranging from 5,000 to 8,000, increasing by increments of 1,000. Based on the experiment in previous work,~\cite{hermelink2021fault}, the 1000 inequalities increments could clearly show the CCA attack time consumption trends while saving the total experiment time. The success rates are shown in \tblref{Percentage error} and the average attack time consumptions are shown in \figref{Time experiment 2}. Instability was observed in Kyber-768 with 6,000 inequalities and in Kyber-1024 with 8,000 inequalities from \tblref{Percentage error} and \figref{Time experiment 2}. These results suggest that to achieve stability, Kyber-512 requires more than 5,000 inequalities, Kyber-768 requires more than 6,000, and Kyber-1024 requires more than 8,000.

\begin{table}[htbp]
\centering
\caption{Fault-enabled CCA Success Rate}
\tbllbl{Percentage error} 
\begin{scriptsize}
\begin{tabular}{|c|c|c|c|c|c|c|c|}
\hline
\textbf{Kyber Version} & \textbf{Iteration} & \textbf{Threads} & \textbf{5000} & \textbf{6000} & \textbf{7000} & \textbf{8000} \\ \hline
512                    & 100                & 64              & 94.04\%       & 91.19\%       & 85.16\%       & 91.39\%       \\ \hline
768                    & 100                & 64              & 93.71\%       & \textbf{34.81\%} & 95.08\%    & 94.09\%       \\ \hline
1024                   & 100                & 64              & 99.54\%       & 99.50\%       & 99.57\%       & \textbf{79.24\%} \\ \hline
\end{tabular}
\end{scriptsize}
\end{table}

\subsection{Shortest Time Cost and Proper Inequality Number Range}
 The time cost for each fault attack was recorded throughout the experiment to identify the minimum time required for successful attacks. Based on the results from previous experiments, we determined the optimal range of inequalities for each Kyber level, balancing the likelihood of a successful attack while minimizing time consumption. The results are summarized in \tblref{Shortest time cost and inequality number range}. The shortest time recorded for each Kyber level was as follows: Kyber-512 took 183 seconds with 7000 inequalities, Kyber-768 took 337 seconds with 8000 inequalities, and Kyber-1024 took 615 seconds with 11,000 inequalities.

 \begin{table}[htbp]
\centering
    \caption{Best time cost and inequality number range}
    \begin{tabular}{cccccc}
    \hline
Kyber level & Best time Cost (s) & Inequality   & Inequality  range \\ \hline
Kyber-512   & 183                 & 7000               & 5000-8000 \\ \hline
Kyber-768   & 337                 & 8000               & 6000-10000 \\ \hline
Kyber-1024  & 615                 & 11000              & 8000-13000 \\ \hline
    \end{tabular}
    \tbllbl{Shortest time cost and inequality number range}
\end{table}

\section{Conclusions}



This study presented a general approach to encrypting ITS communications using Kyber KEM by mapping Kyber KEM's different versions to the different security levels of ITS communications. In addition, we assessed Kyber's vulnerabilities to a hardware-based attack, such as a fault-enabled CCA. Our investigations found that the different security levels associated with the different Kyber versions (i.e., low, medium, and high-security levels) require varying numbers of inequalities to reveal the secret key in Kyber-based communication. These security levels also directly affect the time needed to perform a successful attack. For example, Kyber-1024 was found to offer the strongest protection (as it required the maximum time for the fault-enabled CCA to reveal the corresponding secret key) among the three versions of Kyber considered in this study, i.e., Kyber-512, Kyber-768, and Kyber-1024. These findings underscore the importance of carefully selecting parameters to balance security and performance in practical cryptographic applications.

The approach presented in this study is a generalized approach to securing any ITS communications with a standardized PQC algorithm, Kyber. Although the use case considered in this study is V2X communication in an ITS application, this approach can be transferred to other domains, such as healthcare, government, legal and financial services, which require strong encryption protection. Consequently, the vulnerabilities of Kyber explored using a fault-enabled CCA in this study could also affect Kyber's applicability to the other domains. Protection from such hardware-based vulnerabilities requires further exploration.

This study primarily focused on evaluating hardware-based fault injection attacks on Kyber. However, this work does not include other potential vulnerabilities, such as side-channel attacks. Future research could extend the analysis to include more comprehensive attack types to evaluate Kyber's resilience across different threat models in ITS-related communications.

\section*{Declarations }
\subsection*{Availability of Data and Materials}
The data that support the findings of this study are available from the corresponding author, K.Z., upon reasonable request. 


\subsection*{Funding}
This work is partially supported by the National Center for Transportation Cybersecurity and Resiliency (TraCR) (a U.S. Department of Transportation National University Transportation Center) headquartered at Clemson University, Clemson, South Carolina, USA.


\subsection*{Acknowledgment}
This work is partially supported by the National Center for Transportation Cybersecurity and Resiliency (TraCR) (a U.S. Department of Transportation National University Transportation Center) headquartered at Clemson University, Clemson, South Carolina, USA. Any opinions, findings, conclusions, and recommendations expressed in this material are those of the author(s) and do not necessarily reflect the views of TraCR, and the U.S. Government assumes no liability for the contents or use thereof.


\begin{thebibliography}{29}
\ifx \bisbn   \undefined \def \bisbn  #1{ISBN #1}\fi
\ifx \binits  \undefined \def \binits#1{#1}\fi
\ifx \bauthor  \undefined \def \bauthor#1{#1}\fi
\ifx \batitle  \undefined \def \batitle#1{#1}\fi
\ifx \bjtitle  \undefined \def \bjtitle#1{#1}\fi
\ifx \bvolume  \undefined \def \bvolume#1{\textbf{#1}}\fi
\ifx \byear  \undefined \def \byear#1{#1}\fi
\ifx \bissue  \undefined \def \bissue#1{#1}\fi
\ifx \bfpage  \undefined \def \bfpage#1{#1}\fi
\ifx \blpage  \undefined \def \blpage #1{#1}\fi
\ifx \burl  \undefined \def \burl#1{\textsf{#1}}\fi
\ifx \doiurl  \undefined \def \doiurl#1{\url{https://doi.org/#1}}\fi
\ifx \betal  \undefined \def \betal{\textit{et al.}}\fi
\ifx \binstitute  \undefined \def \binstitute#1{#1}\fi
\ifx \binstitutionaled  \undefined \def \binstitutionaled#1{#1}\fi
\ifx \bctitle  \undefined \def \bctitle#1{#1}\fi
\ifx \beditor  \undefined \def \beditor#1{#1}\fi
\ifx \bpublisher  \undefined \def \bpublisher#1{#1}\fi
\ifx \bbtitle  \undefined \def \bbtitle#1{#1}\fi
\ifx \bedition  \undefined \def \bedition#1{#1}\fi
\ifx \bseriesno  \undefined \def \bseriesno#1{#1}\fi
\ifx \blocation  \undefined \def \blocation#1{#1}\fi
\ifx \bsertitle  \undefined \def \bsertitle#1{#1}\fi
\ifx \bsnm \undefined \def \bsnm#1{#1}\fi
\ifx \bsuffix \undefined \def \bsuffix#1{#1}\fi
\ifx \bparticle \undefined \def \bparticle#1{#1}\fi
\ifx \barticle \undefined \def \barticle#1{#1}\fi
\bibcommenthead
\ifx \bconfdate \undefined \def \bconfdate #1{#1}\fi
\ifx \botherref \undefined \def \botherref #1{#1}\fi
\ifx \url \undefined \def \url#1{\textsf{#1}}\fi
\ifx \bchapter \undefined \def \bchapter#1{#1}\fi
\ifx \bbook \undefined \def \bbook#1{#1}\fi
\ifx \bcomment \undefined \def \bcomment#1{#1}\fi
\ifx \oauthor \undefined \def \oauthor#1{#1}\fi
\ifx \citeauthoryear \undefined \def \citeauthoryear#1{#1}\fi
\ifx \endbibitem  \undefined \def \endbibitem {}\fi
\ifx \bconflocation  \undefined \def \bconflocation#1{#1}\fi
\ifx \arxivurl  \undefined \def \arxivurl#1{\textsf{#1}}\fi
\csname PreBibitemsHook\endcsname

\bibitem[\protect\citeauthoryear{}{}]{key}
\begin{botherref}
\url{https://www.transportation.gov/v2x}.
[Accessed 28-09-2024]
\end{botherref}
\endbibitem

\bibitem[\protect\citeauthoryear{}{}]{qualcommCV2XAuto}
\begin{botherref}
{C}-{V}2{X} {A}uto {T}echnology | {T}he {F}uture of {A}utonomous {C}onnectivity|{Q}ualcomm qualcomm.com.
\url{https://www.qualcomm.com/products/automotive/c-v2x/overview}.
[Accessed 28-09-2024]
\end{botherref}
\endbibitem

\bibitem[\protect\citeauthoryear{{ARC-IT}}{2024}]{ARC-IT}
\begin{botherref}
\oauthor{\bsnm{{ARC-IT}}}:
Architecture Reference for Cooperative and Intelligent Transportation.
\url{https://www.arc-it.net/index.htm}
(2024)
\end{botherref}
\endbibitem

\bibitem[\protect\citeauthoryear{{National Institute of Standards and Technology (NIST)}}{2020}]{round3}
\begin{botherref}
\oauthor{\bsnm{{National Institute of Standards and Technology (NIST)}}}:
Post-Quantum Cryptography Standardization.
\url{https://csrc.nist.gov/projects/post-quantum-cryptography}
(2020)
\end{botherref}
\endbibitem

\bibitem[\protect\citeauthoryear{{National Institute of Standards and Technology}}{2022}]{round3result}
\begin{botherref}
\oauthor{\bsnm{{National Institute of Standards and Technology}}}:
Post-Quantum Cryptography Standardization.
\url{https://csrc.nist.gov/Projects/post-quantum-cryptography/selected-algorithms-2022}
(2022)
\end{botherref}
\endbibitem

\bibitem[\protect\citeauthoryear{}{}]{key2}
\begin{botherref}
\url{https://www.nist.gov/news-events/news/2024/08/nist-releases- first-3-finalized-post-quantum-encryption-standards}.
[Accessed 30-09-2024]
\end{botherref}
\endbibitem

\bibitem[\protect\citeauthoryear{Laboratory}{2023}]{nistfips203}
\begin{botherref}
\oauthor{\bsnm{Laboratory}, \binits{I.T.}}:
{FIPS} 203 module-lattice-based key-encapsulation mechanism standard.
Technical report,
National Institute of Standards and Technology
(2023)
\end{botherref}
\endbibitem

\bibitem[\protect\citeauthoryear{}{2023}]{nistfips204}
\begin{botherref}
{FIPS} 204 module-lattice-based digital signature standard.
Technical report,
National Institute of Standards and Technology
(2023)
\end{botherref}
\endbibitem

\bibitem[\protect\citeauthoryear{Bao et~al.}{2024}]{bao2024aeka}
\begin{barticle}
\bauthor{\bsnm{Bao}, \binits{T.}},
\bauthor{\bsnm{He}, \binits{P.}},
\bauthor{\bsnm{Xie}, \binits{J.}},
\bauthor{\bsnm{Jacinto}, \binits{H.S.}}:
\batitle{Aeka: Fpga implementation of area-efficient karatsuba accelerator for ring-binary-lwe-based lightweight pqc}.
\bjtitle{ACM Transactions on Reconfigurable Technology and Systems}
\bvolume{17}(\bissue{2}),
\bfpage{1}--\blpage{23}
(\byear{2024})
\end{barticle}
\endbibitem

\bibitem[\protect\citeauthoryear{Minhas and Mansoor}{2024}]{minhas2024edge}
\begin{botherref}
\oauthor{\bsnm{Minhas}, \binits{N.N.}},
\oauthor{\bsnm{Mansoor}, \binits{K.}}:
Edge computing-based scheme for post-quantum iot security for e-health.
IEEE Internet of Things Journal
(2024)
\end{botherref}
\endbibitem

\bibitem[\protect\citeauthoryear{}{}]{ibmWorldsFirst}
\begin{botherref}
{W}orld’s first quantum computing safe tape drive --- research.ibm.com.
\url{https://research.ibm.com/blog/crystals-quantum-safe}.
[Accessed 24-07-2024]
\end{botherref}
\endbibitem

\bibitem[\protect\citeauthoryear{Tan et~al.}{2023}]{10323839}
\begin{bchapter}
\bauthor{\bsnm{Tan}, \binits{W.}},
\bauthor{\bsnm{Lao}, \binits{Y.}},
\bauthor{\bsnm{Parhi}, \binits{K.K.}}:
\bctitle{Kybermat: Efficient accelerator for matrix-vector polynomial multiplication in crystals-kyber scheme via ntt and polyphase decomposition}.
In: \bbtitle{2023 IEEE/ACM International Conference on Computer Aided Design (ICCAD)},
pp. \bfpage{1}--\blpage{9}
(\byear{2023}).
\doiurl{10.1109/ICCAD57390.2023.10323839}
\end{bchapter}
\endbibitem

\bibitem[\protect\citeauthoryear{Wang et~al.}{2022}]{wang2022integral}
\begin{bchapter}
\bauthor{\bsnm{Wang}, \binits{A.}},
\bauthor{\bsnm{Tan}, \binits{W.}},
\bauthor{\bsnm{Parhi}, \binits{K.K.}},
\bauthor{\bsnm{Lao}, \binits{Y.}}:
\bctitle{Integral sampler and polynomial multiplication architecture for lattice-based cryptography}.
In: \bbtitle{2022 IEEE International Symposium on Defect and Fault Tolerance in VLSI and Nanotechnology Systems (DFT)},
pp. \bfpage{1}--\blpage{6}
(\byear{2022}).
\bcomment{IEEE}
\end{bchapter}
\endbibitem

\bibitem[\protect\citeauthoryear{Shahmirzadi and Moradi}{2020}]{9250822}
\begin{bchapter}
\bauthor{\bsnm{Shahmirzadi}, \binits{A.R.}},
\bauthor{\bsnm{Moradi}, \binits{A.}}:
\bctitle{Clock glitch versus sifa}.
In: \bbtitle{2020 IEEE International Symposium on Defect and Fault Tolerance in VLSI and Nanotechnology Systems (DFT)},
pp. \bfpage{1}--\blpage{6}
(\byear{2020}).
\doiurl{10.1109/DFT50435.2020.9250822}
\end{bchapter}
\endbibitem

\bibitem[\protect\citeauthoryear{Ravi et~al.}{2020}]{ravi2020drop}
\begin{barticle}
\bauthor{\bsnm{Ravi}, \binits{P.}},
\bauthor{\bsnm{Bhasin}, \binits{f.}},
\bauthor{\bsnm{Roy}, \binits{S.S.}},
\bauthor{\bsnm{Chattopadhyay}, \binits{A.}}:
\batitle{Drop by drop you break the rock-exploiting generic vulnerabilities in lattice-based pke/kems using em-based physical attacks.}
\bjtitle{IACR Cryptol. ePrint Arch.}
\bvolume{2020},
\bfpage{549}
(\byear{2020})
\end{barticle}
\endbibitem

\bibitem[\protect\citeauthoryear{Guo et~al.}{2020}]{guo2020key}
\begin{bchapter}
\bauthor{\bsnm{Guo}, \binits{Q.}},
\bauthor{\bsnm{Johansson}, \binits{T.}},
\bauthor{\bsnm{Nilsson}, \binits{A.}}:
\bctitle{A key-recovery timing attack on post-quantum primitives using the fujisaki-okamoto transformation and its application on frodokem}.
In: \bbtitle{Annual International Cryptology Conference},
pp. \bfpage{359}--\blpage{386}
(\byear{2020}).
\bcomment{Springer}
\end{bchapter}
\endbibitem

\bibitem[\protect\citeauthoryear{Bhasin et~al.}{2021}]{bhasin2021attacking}
\begin{botherref}
\oauthor{\bsnm{Bhasin}, \binits{S.}},
\oauthor{\bsnm{D’Anvers}, \binits{J.-P.}},
\oauthor{\bsnm{Heinz}, \binits{D.}},
\oauthor{\bsnm{P{\"o}ppelmann}, \binits{T.}},
\oauthor{\bsnm{Van~Beirendonck}, \binits{M.}}:
Attacking and defending masked polynomial comparison for lattice-based cryptography.
IACR Transactions on Cryptographic Hardware and Embedded Systems,
334--359
(2021)
\end{botherref}
\endbibitem

\bibitem[\protect\citeauthoryear{Oder et~al.}{2016}]{oder2016practical}
\begin{botherref}
\oauthor{\bsnm{Oder}, \binits{T.}},
\oauthor{\bsnm{Schneider}, \binits{T.}},
\oauthor{\bsnm{P{\"o}ppelmann}, \binits{T.}},
\oauthor{\bsnm{G{\"u}neysu}, \binits{T.}}:
Practical cca2-secure and masked ring-lwe implementation.
Cryptology ePrint Archive
(2016)
\end{botherref}
\endbibitem

\bibitem[\protect\citeauthoryear{Hamburg et~al.}{2021}]{hamburg2021chosen}
\begin{botherref}
\oauthor{\bsnm{Hamburg}, \binits{M.}},
\oauthor{\bsnm{Hermelink}, \binits{J.}},
\oauthor{\bsnm{Primas}, \binits{R.}},
\oauthor{\bsnm{Samardjiska}, \binits{S.}},
\oauthor{\bsnm{Schamberger}, \binits{T.}},
\oauthor{\bsnm{Streit}, \binits{S.}},
\oauthor{\bsnm{Strieder}, \binits{E.}},
\oauthor{\bsnm{Vredendaal}, \binits{C.}}:
Chosen ciphertext k-trace attacks on masked cca2 secure kyber.
IACR Transactions on Cryptographic Hardware and Embedded Systems,
88--113
(2021)
\end{botherref}
\endbibitem

\bibitem[\protect\citeauthoryear{Fujisaki and Okamoto}{1999}]{fujisaki1999secure}
\begin{bchapter}
\bauthor{\bsnm{Fujisaki}, \binits{E.}},
\bauthor{\bsnm{Okamoto}, \binits{T.}}:
\bctitle{Secure integration of asymmetric and symmetric encryption schemes}.
In: \bbtitle{Annual International Cryptology Conference},
pp. \bfpage{537}--\blpage{554}
(\byear{1999}).
\bcomment{Springer}
\end{bchapter}
\endbibitem

\bibitem[\protect\citeauthoryear{Hofheinz et~al.}{2017}]{hofheinz2017modular}
\begin{bchapter}
\bauthor{\bsnm{Hofheinz}, \binits{D.}},
\bauthor{\bsnm{H{\"o}velmanns}, \binits{K.}},
\bauthor{\bsnm{Kiltz}, \binits{E.}}:
\bctitle{A modular analysis of the fujisaki-okamoto transformation}.
In: \bbtitle{Theory of Cryptography Conference},
pp. \bfpage{341}--\blpage{371}
(\byear{2017}).
\bcomment{Springer}
\end{bchapter}
\endbibitem

\bibitem[\protect\citeauthoryear{Hermelink et~al.}{2021}]{hermelink2021fault}
\begin{bchapter}
\bauthor{\bsnm{Hermelink}, \binits{J.}},
\bauthor{\bsnm{Pessl}, \binits{P.}},
\bauthor{\bsnm{P{\"o}ppelmann}, \binits{T.}}:
\bctitle{Fault-enabled chosen-ciphertext attacks on kyber}.
In: \bbtitle{Progress in Cryptology--INDOCRYPT 2021: 22nd International Conference on Cryptology in India, Jaipur, India, December 12--15, 2021, Proceedings 22},
pp. \bfpage{311}--\blpage{334}
(\byear{2021}).
\bcomment{Springer}
\end{bchapter}
\endbibitem

\bibitem[\protect\citeauthoryear{Prokop and Pe{\ss}l}{2021}]{prokop2021fault}
\begin{barticle}
\bauthor{\bsnm{Prokop}, \binits{L.}},
\bauthor{\bsnm{Pe{\ss}l}, \binits{P.}}:
\batitle{Fault attacks on cca-secure lattice kems}.
\bjtitle{IACR Transactions on Cryptographic Hardware and Embedded Systems}
\bvolume{2021}(\bissue{2}),
\bfpage{37}--\blpage{60}
(\byear{2021})
\end{barticle}
\endbibitem

\bibitem[\protect\citeauthoryear{}{}]{cryptopediaKyberDoes}
\begin{botherref}
{K}yber - {H}ow does it work? | {A}pproachable {C}ryptography --- cryptopedia.dev.
\url{https://cryptopedia.dev/posts/kyber/}.
[Accessed 28-09-2024]
\end{botherref}
\endbibitem

\bibitem[\protect\citeauthoryear{Bos et~al.}{2018}]{bos2018crystals}
\begin{bchapter}
\bauthor{\bsnm{Bos}, \binits{J.}},
\bauthor{\bsnm{Ducas}, \binits{L.}},
\bauthor{\bsnm{Kiltz}, \binits{E.}},
\bauthor{\bsnm{Lepoint}, \binits{T.}},
\bauthor{\bsnm{Lyubashevsky}, \binits{V.}},
\bauthor{\bsnm{Schanck}, \binits{J.M.}},
\bauthor{\bsnm{Schwabe}, \binits{P.}},
\bauthor{\bsnm{Seiler}, \binits{G.}},
\bauthor{\bsnm{Stehl{\'e}}, \binits{D.}}:
\bctitle{Crystals-kyber: a cca-secure module-lattice-based kem}.
In: \bbtitle{2018 IEEE European Symposium on Security and Privacy (EuroS\&P)},
pp. \bfpage{353}--\blpage{367}
(\byear{2018}).
\bcomment{IEEE}
\end{bchapter}
\endbibitem

\bibitem[\protect\citeauthoryear{}{}]{j2735}
\begin{botherref}
\url{https://www.sae.org/standards/content/j2735_202409/}
\end{botherref}
\endbibitem

\bibitem[\protect\citeauthoryear{Karim and Rawat}{2022}]{9344812}
\begin{barticle}
\bauthor{\bsnm{Karim}, \binits{H.}},
\bauthor{\bsnm{Rawat}, \binits{D.B.}}:
\batitle{Tollsonly please—homomorphic encryption for toll transponder privacy in internet of vehicles}.
\bjtitle{IEEE Internet of Things Journal}
\bvolume{9}(\bissue{4}),
\bfpage{2627}--\blpage{2636}
(\byear{2022})
\doiurl{10.1109/JIOT.2021.3056240}
\end{barticle}
\endbibitem

\bibitem[\protect\citeauthoryear{Deng et~al.}{2020}]{deng2020location}
\begin{barticle}
\bauthor{\bsnm{Deng}, \binits{X.}},
\bauthor{\bsnm{Xin}, \binits{X.}},
\bauthor{\bsnm{Gao}, \binits{T.}}:
\batitle{A location privacy protection scheme based on random encryption period for vsns}.
\bjtitle{Journal of Ambient Intelligence and Humanized Computing}
\bvolume{11},
\bfpage{1351}--\blpage{1359}
(\byear{2020})
\end{barticle}
\endbibitem

\bibitem[\protect\citeauthoryear{}{2019}]{BSM}
\begin{botherref}
Tampa CV Pilot Basic Safety Message (BSM).
\url{http://doi.org/10.21949/1504502/}.
Tampa Hillsborough Expressway Authority
(2019)
\end{botherref}
\endbibitem

\end{thebibliography}

\end{document}